\documentclass[aps,showpacs,showkeys,groupedaddress]{revtex4}

\usepackage{epsfig}
\usepackage{epsf}

\def\slash#1{\setbox0=\hbox{$#1$}
   \dimen0=\wd0 \setbox1=\hbox{/} \dimen1=\wd1
   \ifdim\dimen0>\dimen1 \rlap{\hbox to \dimen0{\hfil/\hfil}} #1
   \else  \rlap{\hbox to \dimen1{\hfil$#1$\hfil}} / \fi}
   
\def \intk{\int \frac{d^2 k}{(2\pi)^2}}

\def\intdw{\int \frac{d^{2\omega} k}{(2\pi)^{2\omega}}}

\def\intkum{\int_{-\Lambda_1}^{\Lambda_1} \frac{dk_1}{2\pi}}
\def\intSC{\intkum \int_{-\Lambda_0}^{\Lambda_0} \frac{dk_0}{2\pi}}

\begin{document}

\title{Momentum Space Regularizations and the Indeterminacy in the Schwinger Model}%

\author{C. W. Morais}
\author{A. L. Mota}
\email{motaal@ufsj.edu.br}
\affiliation{Departamento de Ci\^{e}ncias Naturais, Universidade Federal de S\~{a}o Jo\~{a}o del Rei,
C.P. 110,  CEP 36301-160, S\~ao Jo\~ao del Rei, Brazil}

\begin{abstract}
We revisited the problem of the presence of finite indeterminacies that appear in the calculations of a 
Quantum Field Theory. We investigate the occurrence of undetermined mathematical quantities in the evaluation of the Schwinger model
in several regularization scenarios. We show that the undetermined character of the divergent part 
of the vacuum polarization tensor of the model, introduced as an {\it ansatz} in previous works, can be obtained
mathematically if one introduces a set of two parameters in the evaluation of these quantities. The formal
mathematical properties of this tensor and their violations are discussed. The analysis
is carried out in both analytical and sharp cutoff regularization procedures. We also show how the Pauli Villars
regularization scheme eliminates the indeterminacy, giving a gauge invariant result in the vector Schwinger model. 
\end{abstract}


\maketitle

\section{Introduction}

The Schwinger model \cite{Schwinger:1962tp}, Quantum Electrodynamics (QED) in two dimensions with massless fermions, 
is the simplest model for fermionic fields
that implements several features observed also in other 4D more realistic gauge models.
It is an exactly solvable model \cite{Manton:1985jm} that exhibits gauge invariance in its vector version and also
the presence of the axial anomaly in the chiral version.  In this model, the photon
receives a finite dynamical mass that can be evaluated exactly to all orders in the perturbative expansion of the quantum action. 
It is also an interesting feature of the model that, being two dimensional, the vector and axial vector fermionic currents are not independent of each other.
So, the vacuum polarization tensor of the chiral Schwinger model is related with the same quantity computed in the
vector Schwinger model \cite{Jackiw:1984zi}. Finally, although divergences appear in the intermediate calculations within the model, the quantum action of the Schwinger model is finite. Nevertheless, 
the vacuum polarization tensor has different results in different regularization schemes, being recognized, for this
reason, as an undetermined quantity\cite{Jackiw:1999qq}. 

Gauge invariance can be obtained within the model by adopting the {\it ansatz } that this divergent part of the vacuum polarization tensor is a undetermined quantity
that appears in the final result as an ambiguity to be fixed by some symmetry relation \cite{Jackiw:1984zi}.
In this process, an apparent formal characteristic - the tracelessness of the vacuum polarization tensor - has
to be abandoned. While from the mathematical point of view one should expect this null trace feature of
the vacuum polarization tensor to be preserved up to the end of the calculations, it is clear that its lost is due to the fact that, although presenting a finite final result, this amplitude is related to the product of local operators and, thus, not well defined. A (finite) renormalization of the model by the introduction of invariant counter terms \cite{Bonneau:2000ai,Bonneau:2006ma} is necessary. Nevertheless, when applying a regularization scheme
to deal with the intermediate divergences of the model one should expect that the formal mathematical properties of the
amplitudes would be preserved, and it is natural to ask how this null trace property is lost in the regularization procedure 
and whether the regularizations schemes are capable to reproduce the undetermined character of the ambiguity.
The aim of this paper is to answer these two questions in the framework of different regularization schemes, showing that the ambiguity appears as a mathematical indeterminacy when the regularization is carried out with a set of two parameters, and to discuss its implications to the Schwinger and chiral Schwinger models.

In the chiral Schwinger model, it was shown that it is not possible to fix the ambiguity in order to
obtain the conservation of the chiral current. This is the two dimensional version of the axial anomaly problem \cite{Adler:1969gk,Bell:1969ts},
where the axial-vector-vector three point function presents a finite undetermined part that cannot be fixed in order to
satisfy simultaneously the gauge and chiral Ward identities \cite{Jackiw:1999qq,BaetaScarpelli:2001ix}. 
The axial anomaly is an important and well established problem where the presence of finite undetermined quantities 
plays a crucial role in the selection of which symmetry relation has to be violated by the anomaly - in t'Hoofts proton decay 
calculation \cite{tHooft:1976up}, the ambiguity is fixed in order to preserve chiral symmetry, violating gauge invariance, whereas, in
QCD, the ambiguity is fixed in such a way that the gauge invariance is satisfied, with the cost of violating chiral symmetry \cite{Adler:1969gk,Bell:1969ts}. 
No one is able to fix {\it a priori} the result for the ambiguity, and to do so at the end of the calculations, some
physical relation has to be employed.
For example, in the context of an effective chiral 
quark model of QCD, it was shown recently \cite{Ferreira:2008tu} that the model can be made free from this ambiguity by
demanding the finiteness of the axial-vector coupling, leading to the expected violation of the chiral Ward identity
for QCD. It is therefore interesting to be able to employ regularization schemes that allow the identification of the potentially ambiguous quantities, preserving the indeterminacy to be fixed by any initial choice.
Nevertheless, only in a relatively recent past some new regularization 
schemes \cite{Freedman:1991tk,BaetaScarpelli:2001ix}, in both configuration and momentum spaces, incorporate the possibility of
keeping these quantities as undetermined up to the very end of calculations, with symmetry relations being used to fix
their values.

This kind of indeterminacies appears also in several other situations in
Quantum Field Theory (QFT), as for example in the recent investigations on the radiatively generation of Lorentz and CPT
violating Chern-Simons term in QED \cite{Jackiw:1999yp,Scarpelli:2008fw}, in the study of 2D gravitational anomalies for Weyl massless
fermions in a gravitational background \cite{AlvarezGaume:1983ig,Langouche:1984sy,Leutwyler:1985ar,Souza:2005vf}, 
in phenomenological applications to the linear sigma model \cite{Dias:2005tb,Hiller:2005uw}, in the
study of the Thirring-Wess model \cite{Thirring:1964,Casana:2004nv} and so on. 
It also stands as an open possibility in the coupling of the electromagnetic field to other 2D fermionic systems, due their similarities with the Schwinger model, as in the coupling of the 2D Gross Neveu model to gauge fields 
\cite{Gross:1974jv,Tracas:1990kw,Franzki:1995fn} and its applications on the study of the polyacetylene \cite{Takayama:1980zz,Caldas:2008zz,Caldas:2008xa}.

In the configuration space, the problem of evaluating the product of local operators can be properly addressed by 
placing the operators at distinct points joined by the introduction of a Wilson line \cite{Peskin:1995ev}. 
However, in some cases it could be more appropriate to carry out this evaluation in the momentum space, for example,
for the calculations of amplitudes with fixed external momenta. 
The so called Implicit Regularization procedure \cite{BaetaScarpelli:2001ix} is carried out in momentum space, but the 
ambiguous quantities are to be recognized by prior knowledge of their results in different regularizations. Also, the
indeterminacies are parametrized, as an {\it ansatz}, in the spirit of the ideas introduced in Ref. \cite{Jackiw:1999qq}.
So, how the indeterminacies can be obtained from the mathematical point of view in momentum space regularization procedures
is an important issue to be analyzed in order to allow their identification and to justify the parametrizations employed in the Implicit
Regularization procedure.
In this sense, the Schwinger model presents the perfect {\it scenario} to this study, since all the
relevant amplitudes to be computed are finite, but present an indeterminacy, and the two dimensionality of the problem simplifies the computation
of the surface terms present in the evaluation of the ambiguity. Besides, the generalization of the
results to higher dimensions is straightforward. 

In this paper we will evaluate the vacuum polarization tensor in the Schwinger model employing the most popular momentum
space regularization procedures. We will show that the evaluation of this amplitude in terms of two independent parameters 
shows that it presents an indeterminacy, in the mathematical sense. To raise the indeterminacy one should be able to choose
a specific path in order to evaluate connection limits involved, and only physical (symmetry) arguments
can indicate the correct choice, in the spirit of Refs. \cite{Jackiw:1999qq,Jackiw:2000di}. 

This paper is organized as follows: in section II we review the evaluation of the vacuum polarization tensor in the vector
Schwinger model and isolate the ambiguous integral related to it. In section III we compute the ambiguity in analytical
regularization schemes and show that this corresponds to a mathematical indeterminacy. We discuss the limits where 
gauge invariance is obtained and why the tracelessness property of the tensor is lost. In section IV we address the same
problem in the context of cutoff regularizations, in both Euclidean and Minkowski spaces. Section V brings an analysis
of the problem in the context of the Pauli Villars regularization scheme, and we discuss why there is no ambiguity in
this regularization. In section VI we discuss the implications of the results obtained to the chiral Schwinger model.
Finally, section VII brings our conclusions.

\section{The Model}

The Schwinger model is defined by the following Lagrangian
\begin{equation}
L=i\bar{\psi}\partial\psi-e\bar{\psi}\gamma^{\mu}A_{\mu}\psi - \frac{1}{4}(F_{\mu\nu})^2, \label{Lag}
\end{equation}
with $\{\gamma^\mu, \gamma^\nu\}=2g^{\mu\nu}$.
The vacuum polarization tensor is evaluated from the quantum action computed with the Lagrangian (\ref{Lag}) in the usual way, resulting in
the amplitude
\begin{equation}
\Pi^{\mu\nu}(p)=ie^2 \intk Tr \Big\{ \gamma^{\mu} \frac{\slash{k}}{k^2+i\epsilon} \gamma^{\nu} \frac{(\slash{k}+\slash{p})}{(k+p)^2+i\epsilon} \Big\}. \label{vacpol0}
\end{equation}

Eq. (\ref{vacpol0}) is superficially logarithmically divergent. In exact two dimension, one should expect $\Pi^{\mu\nu}$ to be traceless, 
i.e., $\Pi^{\mu}_{\mu}(p)=0$, since, in this case, $\gamma^{\mu}\gamma^{\sigma}\gamma_{\mu}=0$. Taking the Dirac traces in Eq.
(\ref{vacpol0}) and introducing one Feynman parameter, keeping the $i\epsilon$ prescription implicit, we get
\begin{equation}
\Pi^{\mu\nu}(p)=S^{\mu\nu} + \Pi_{fin}^{\mu\nu}(p), \label{vacpol}
\end{equation}
with
\begin{equation}
S^{\mu\nu}= 2ie^2 \int_0^1 dx \intk \frac{2k^{\mu}k^{\nu}-g^{\mu\nu}k^2}{(k^2-M^2)^2} \label{PiInf}
\end{equation}
and
\begin{equation}
\Pi_{fin}^{\mu\nu}(p)= -2ie^2 (2p^\mu p^\nu - g^{\mu\nu}p^2) \intk \frac{x(1-x)} {(k^2-M^2)^2}.
\end{equation}
with $M^2=p^2 x^2-p^2 x$. 

The integral $\Pi_{fin}^{\mu\nu}(p)$ is convergent, and, after applying a regularization
scheme and taking the connection limit, one gets
\begin{equation}
\Pi_{fin}^{\mu\nu}(p) = \frac{e^2}{\pi}\Big( \frac{g^{\mu\nu}}{2} - \frac{p^{\mu}p^{\nu}}{p^2} \Big). \label{finite}
\end{equation}

It is interesting to note that $S^{\mu\nu}$ does not depend on $M^2$. It can
be verified by applying the following identity in Eq.(\ref{PiInf}):
\begin{equation}
\frac{1}{(k^2-M^2)^2}=\frac{1}{(k^2-\mu^2)^2}+\frac{2(M^2-\mu^2)}{(k^2-\mu^2)^2(k^2-M^2)}+\frac{(M^2-\mu^2)^2}{(k^2-\mu^2)^2(k^2-M^2)^2}
\end{equation}
with $\mu^2$ arbitrary. After this, one gets
\begin{eqnarray}
&& S^{\mu\nu} = 2ie^2 \intk \frac{k^{\mu}k^{\nu}-g^{\mu \nu}k^2}{(k^2-\mu^2)^2} \label{PiInfMu} \\
&& + 2ie^2 \int_0^1 dx \intk (k^{\mu}k^{\nu}-g^{\mu \nu}k^2) \Big( \frac{2(M^2-\mu^2)}{(k^2-\mu^2)^2(k^2-M^2)}+\frac{(M^2-\mu^2)^2}{(k^2-\mu^2)^2(k^2-M^2)^2} \Big) \nonumber
\end{eqnarray}
The last integral on the right hand side of Eq.(\ref{PiInfMu}) are finite and can be computed within any regularization scheme,
giving a vanishing result. Thus, we get
\begin{equation}
S^{\mu\nu} = 2ie^2 \intk \frac{2k^{\mu}k^{\nu}-g^{\mu\nu}k^2}{(k^2-\mu^2)^2}. \label{PiInfFin}
\end{equation}
As it is well known, $S^{\mu\nu}$ has a regularization dependent result \cite{BaetaScarpelli:2000zs}. As $\mu^2$ is arbitrary, $S^{\mu\nu}$ cannot depend on $\mu^2$. 
The identification of $S^{\mu\nu}$ with a surface term follows
directly, since
\begin{equation}
S^{\mu\nu}|_{\mu^2=0} = \intk \frac{2k^\mu k^\nu - g^{\mu\nu}k^2}{(k^2)^2} = -\intk \frac{\partial}{\partial k^\mu} \Big\{ \frac{k^\nu}{k^2} \Big\}.
\end{equation}

The action of the model presents gauge invariance. Physically, this feature must be verified to all orders in the
perturbative expansion of the action, no anomalies are expected in the vector Schwinger model, since it has no
parity violating terms. It is interesting to observe, however, that it is not possible to preserve simultaneously
the gauge invariance and the tracelessness of the vacuum polarization tensor. One can see this by
writing the most general form of the divergent part of the vacuum polarization tensor compatible with its second-hank
tensorial structure and with the {\it a priori} dependence with the external momentum $p$
\begin{equation}
S^{\mu\nu} = \frac{e^2}{\pi} \Big(\alpha g^{\mu\nu} + \beta \frac{p^\mu p^\nu}{p^2}\Big). \label{Pimunuparam}
\end{equation}
Replacing Eq.(\ref{Pimunuparam}) and (\ref{finite}) in Eq.(\ref{vacpol}) and imposing both tracelessness ($g_{\mu\nu}\Pi^{\mu\nu}(p)=0$)
and gauge invariance ($p_\mu\Pi^{\mu\nu}(p)=0$), one finds $\alpha=-\frac{1}{2}$ and $\beta=1$, which results in the trivial solution
$S^{\mu\nu} = -\Pi_{fin}^{\mu\nu}(p)$, or $\Pi^{\mu\nu}(p) = 0$. So, one of the formal expected properties
for the vacuum polarization tensor must be abandoned at quantum level. The physical choice is, of course, to maintain gauge
invariance and to allow the violation of the trace identity. It should be stressed that this second feature is not, in fact,
to be expected from the mathematical point of view - the vacuum polarization tensor as expressed by Eq.(\ref{vacpol0}) is
a divergent quantity. Although at exact two dimensions the trace of the term $\gamma^{\mu} \slash{k} \gamma^{\nu}$ appearing
on the integrand of (\ref{vacpol0}) vanishes, the remaining integral diverges, and thus the right hand side of this
equation is a mathematical indeterminacy. In this paper, we will discuss some strategies to correctly display this indeterminacy
in some of the most popular momentum space regularization schemes.

\section{The Vacuum Polarization Tensor in Dimensional Regularization}

In arbitrary $2\omega$ dimensions, the divergent piece of the vacuum polarization tensor of the Schwinger model is given, from Eq.(\ref{PiInfFin}), by
\begin{equation}
S^{\mu\nu} = 2ie^2 \Big( \intdw \frac{2k^{\mu}k^{\nu}}{(k^2-\mu^2)^2} - \intdw \frac{g^{\mu\nu}k^2}{(k^2-\mu^2)^2} \Big). \label{DimReg}
\end{equation}
Solving Eq.(\ref{DimReg}), one gets
\begin{equation}
S^{\mu\nu} = \frac{2e^2g^{\mu\nu}}{(4\pi)^\omega (m^2)^{1-\omega}} \Gamma(2-\omega)
\end{equation}
In two dimensions, with $\omega=1$, we find
\begin{equation}
S^{\mu\nu} = \frac{e^2g^{\mu\nu}}{2\pi} \label{PiInfDR}
\end{equation}
and
\begin{equation}
\Pi^{\mu\nu}(p) = \frac{e^2}{\pi}\Big( g^{\mu\nu} - \frac{p^{\mu}p^{\nu}}{p^2} \Big), \label{PimunuFinal}
\end{equation}
the expected gauge invariant result. As already discussed, due to its divergent character,
one cannot assume that Eq.(\ref{PimunuFinal}) is traceless. 
This can be easily recognized in the dimensional regularization (DR) procedure: in the computation
of Eq.(\ref{vacpol0}) in arbitrary $2\omega$ dimensions, we get $\gamma^{\mu}\gamma^{\sigma}\gamma_{\mu}=2(1-\omega)\gamma^\sigma$. 
This term is null in explicitly 2D, when $\omega=1$. Nevertheless, the remaining integral results proportional
to $\Gamma(1-\omega)$. The product of the two functions is non-null even in the limit $\omega=1$.

The choice of a specific regularization scheme, e.g. the dimensional regularization, selects an {\it a priori} result
to the undetermined integral corresponding to the vacuum polarization amplitude. To see how the undetermined character
of the vacuum polarization tensor can be obtained in the context of analytical regularizations, as dimensional regularization, let us compute
the following quantity
\begin{equation}
S_{\omega\alpha}^{\mu\nu} = 2ie^2 \intdw \frac{2k^{\mu}k^{\nu}-g^{\mu \nu}k^2}{(k^2-\mu^2)^\alpha}, \label{Piomegaalpha}
\end{equation}
that corresponds to $S^{\mu\nu}$ when $\omega=1$ and $\alpha=2$. By performing the dimensional integration,
one finds
\begin{equation}
S_{\omega\alpha}^{\mu\nu} = 2e^2 \frac{g^{\mu\nu}}{(4\pi)^\omega\Gamma(\alpha)(\mu^2)^{\alpha-1-\omega}}(1-\omega)\Gamma(\alpha-1-\omega). \label{Piomegaalpha2}
\end{equation}
For exactly $\alpha=2$ we have $(1-\omega)\Gamma(\alpha-1-\omega)=(1-\omega)\Gamma(1-\omega)=\Gamma(2-\omega)$, resulting in the finite
non-null result given by Eq.(\ref{PiInfDR}) when $\omega=1$. Nevertheless, for $\alpha$ arbitrary this latter property does not apply.
In fact, by replacing $2\omega = 2 - \epsilon$ and $\alpha=2+\delta$ in Eq.(\ref{Piomegaalpha2}), we have
\begin{equation}
S_{\omega\alpha}^{\mu\nu} = 2e^2 \frac{g^{\mu\nu}}{(4\pi)^{1-\epsilon/2}\Gamma(2+\delta)(\mu^2)^{\delta+\epsilon/2}}\frac{\epsilon}{2}\Gamma\Big(\delta+\frac{\epsilon}{2}\Big). \label{Piomegaalpha3}
\end{equation}
As for $\alpha\rightarrow 2$ and $\omega\rightarrow 1$ we have $\delta\rightarrow 0$ and $\epsilon\rightarrow 0$, the Gamma
function appearing on Eq.(\ref{Piomegaalpha3}) can be expanded as usual, resulting in
\begin{equation}
S_{\omega\alpha}^{\mu\nu} = \frac{e^2}{2\pi} g^{\mu\nu} \lim_{\delta,\epsilon \to 0} \frac{\epsilon}{2\delta+\epsilon}. \label{Piomegaalpha4}
\end{equation}
This result displays the indeterminacy in the evaluation of the vacuum polarization tensor in the Schwinger model - it depends
on the path in the $\delta \times \epsilon$ plane employed in order to reach the limit $\delta=0$ and $\epsilon=0$. For instance,
if we choose $\delta=0$ and then take the limit $\epsilon \rightarrow 0$ we reproduce the previous result, Eq.(\ref{PiInfDR}).
Otherwise, by picking $\epsilon = 0$ and taking the limit $\delta \rightarrow 0$, the analytical regularization \cite{Bollini:1972ui}, we 
obtain $S^{\mu\nu} = 0$.

\section{The Vacuum Polarization Tensor in Sharp Cutoff}

The sharp cutoff regularization scheme is well known as being a non gauge invariant procedure. In fact, it is easy
to show that, within this regularization scheme, one gets, by performing the symmetric integration trick
\begin{equation}
\intk k^\mu k^\nu f(k^2) = \frac{g^{\mu\nu}}{2} \intk k^2 f(k^2)
\end{equation}
in the first integral on Eq.(\ref{PiInfFin}) resulting in
\begin{equation}
S^{\mu\nu} = 2ie^2 \intk \frac{2k^{\mu}k^{\nu}-g^{\mu\nu}k^2}{(k^2-\mu^2)^2}=2ie^2 \intk \frac{k^2g^{\mu\nu}-g^{\mu\nu}k^2}{(k^2-\mu^2)^2}=0. \label{Smunueq0}
\end{equation}
Thus, one obtains, from Eqs.(\ref{vacpol}) and (\ref{finite}),
\begin{equation}
\Pi^{\mu\nu}(p) = \frac{e^2}{\pi}\Big( \frac{g^{\mu\nu}}{2} - \frac{p^{\mu}p^{\nu}}{p^2} \Big). \label{vacpolSC}
\end{equation}
It is interesting to note why we do not get the same result in DR (see Eqs.(\ref{PiInfDR}) and (\ref{PimunuFinal})) - 
the application of the symmetric integration trick implies in
\begin{equation}
\intdw \frac{2k^{\mu}k^{\nu}}{(k^2-\mu^2)^2}= \frac{g^{\mu\nu}}{2\omega} \intdw \frac{k^2}{(k^2-\mu^2)^2},
\end{equation}
and
\begin{equation}
S^{\mu\nu} = 2ie^2 g^{\mu\nu} \Big( \frac{1}{\omega} - 1 \Big) \intdw \frac{k^2}{(k^2-\mu^2)^2}.
\end{equation}
The result would vanish only in exact two dimensions (if one ignores the divergent character of the integral), taking 
the limit $\omega \rightarrow 1$ will give the result displayed in Eq.(\ref{Smunueq0}).

Eq.(\ref{vacpolSC}) preserves the tracelessness of the vacuum polarization tensor, but it is not gauge invariant. However,
if one computes the gauge invariance condition $p_\mu\Pi^{\mu\nu}(p)$ by contracting Eq.(\ref{vacpol0}) directly with $p_\mu$,
we get
\begin{equation}
p_\mu \Pi^{\mu\nu}(p)=ie^2 \intk Tr \Big\{ \slash{p} \frac{\slash{k}}{k^2} \gamma^{\nu} \frac{(\slash{k}+\slash{p})}{(k+p)^2} \Big\} 
= ie^2 \Big\{ \intk \frac{ Tr\{ \slash{k}\gamma^\nu \}}{k^2}- \intk \frac{ Tr\{ (\slash{k}+\slash{p})\gamma^\nu \} }{(k+p)^2} \Big\}. \label{pmuPimunu}
\end{equation}

Shifting $k+p \to k$ in the second integral in the right hand side of Eq.(\ref{pmuPimunu}) will result in $p_\mu \Pi^{\mu\nu}(p)=0$,
as expected. However, this procedure is not allowed in the sharp-cutoff regularization without the proper evaluation of the 
surface term that arises from this operation. Thus, let us extract the surface term
by applying the expansion of the integrand of the second integral on the right hand side of Eq.(\ref{pmuPimunu}), as performed 
in Ref. \cite{Battistel:1998sz}

\begin{eqnarray}
&& \intk \frac{ Tr\{ (\slash{k}+\slash{p})\gamma^\nu \} }{(k+p)^2}=\intk exp\Big( p^\mu \frac{\partial}{\partial k^\mu} \Big) \frac{ Tr\{ \slash{k}\gamma^\nu \} }{k^2} = \nonumber \\
&& \intk \frac{ Tr\{ \slash{k}\gamma^\nu \} }{k^2} + p^\mu \intk  \frac{\partial}{\partial k^\mu} \frac{ Tr\{ \slash{k}\gamma^\nu \} }{k^2} +
 \frac{p^\mu p^\rho}{2!} \intk \frac{\partial^2}{\partial k^\mu \partial k^\rho} \frac{ Tr\{ \slash{k}\gamma^\nu \} }{k^2} + ... \label{expans}
\end{eqnarray}

The first integral in the second line of Eq.(\ref{expans}) cancels out exactly with the first integral in the right hand
side of Eq.(\ref{pmuPimunu}), whereas the last integral in Eq.(\ref{expans}), and all others terms of higher order in $p$, not explicitly displayed, 
are convergent and null. The only remaining integral is 
\begin{equation}
p_\mu \Pi^{\mu\nu}(p)= -ie^2 p_\mu \intk  \frac{\partial}{\partial k^\mu} \frac{ Tr\{ \slash{k}\gamma^\nu \} }{k^2}= -2ie^2 p_\mu \intk \frac{2k^{\mu}k^{\nu}-g^{\mu\nu}k^2}{k^4},
\end{equation}
and, by applying again the symmetric integration trick, we get $p_\mu \Pi^{\mu\nu}(p)=0$. So, paradoxically, the explicit computation
of $\Pi^{\mu\nu}(p)$ in the sharp cutoff regularization scheme gives a non gauge invariant result, Eq.(\ref{vacpolSC}), but
if one computes $p_\mu \Pi^{\mu\nu}(p)$ in the same regularization, by performing the contraction with $p_\mu$ before integrating
in $k$, the gauge invariance condition for the vacuum polarization tensor is verified. This is, of course, a well known result,
and it is due to the fact that $\Pi^{\mu\nu}(p)$ is not well defined. Nevertheless, one should expect the results of a given
regularization scheme to be consistent.


Rather than continue listing all the evils arising from the evaluation of divergent integrals in QFT, let us focus in the specific evaluation
of the undetermined integral, Eq.(\ref{PiInfFin}), in the sharp cutoff regularization scheme. First, we will discuss why the
Wick rotation of this integral to the Euclidean space introduces changes that cannot be neglected in the final result. Then,
we will perform the explicit computation of Eq.(\ref{PiInfFin}) in the Minkowski space and show that the result contains a 
mathematical indeterminacy.

\subsection{Evaluation in Euclidean space}

Our aim, in this subsection, is to evaluate Eq.(\ref{PiInfFin}) by using the sharp cutoff regularization scheme in the Euclidean space. Let us introduce two different 
cutoffs, $\Lambda_0$ and $\Lambda_1$, to the integrals in $k_0$ and $k_1$.
This procedure could produce non covariant results, as we will discuss latter, and, for this reason, we will
explicitly compute $S^{00}$ and $S^{11}$ instead of $S^{\mu\nu}$.
When $\mu=0$ and $\nu=1$ or $\mu=1$ and $\nu=0$, the result for Eq.(\ref{PiInfFin}) vanishes trivially. For $\mu=0$ and $\nu=0$, we obtain
\begin{equation}
S_{reg}^{00} = 2ie^2 \Big\{ \intSC \frac{2k_0^2}{(k_0^2-k_1^2-\mu^2+i\epsilon)^2} - \intSC \frac{k_0^2-k_1^2}{(k_0^2-k_1^2-\mu^2+i\epsilon)^2} \Big\},
\end{equation}
where the subscript {\it reg} stands for the sharp cutoff regularized divergent part of the vacuum polarization tensor and
we show explicitly the use of $i\epsilon$ prescription, for a while. A similar expression, with the appropriated interchange
between $k_0$ and $k_1$, is obtained for $\Pi_{reg}^{11}$. Simplifying the expression above, one obtains
\begin{equation}
S_{reg}^{00} = 2ie^2 \Big\{ \intSC \frac{k_0^2}{(k_0^2-k_1^2-\mu^2+i\epsilon)^2} + \intSC \frac{k_1^2}{(k_0^2-k_1^2-\mu^2+i\epsilon)^2} \Big\}. \label{Wick}
\end{equation}

It is a well known fact that due the divergent character of the integral, the Wick rotation cannot be justified. When regulating
the integrals via the finite cutoff procedure, the result of the contour integrals should be took into account, resulting in
different functions of $\Lambda_0$ for the two integrals of Eq.(\ref{Wick}), since the integrands are different functions of $k_0$.
The connection limits $\Lambda_0 \to \infty$ and $\Lambda_1 \to \infty$ should be carried out only at the very end of the computation,
and these contour integrals cannot be neglected. Of course, if one naively neglect the contour contributions to Eq.(\ref{Wick}),
performing the Wick rotation, one gets
\begin{equation}
S_{reg}^{00} = 2e^2 \Big\{ \intSC \frac{k_0^2}{(k_0^2+k_1^2+\mu^2-i\epsilon)^2} - \intSC \frac{k_1^2}{(k_0^2+k_1^2+\mu^2-i\epsilon)^2} \Big\},
\end{equation}
and a simple interchange of variables $k_0 \leftrightarrow k_1$ on the second integral, with $\Lambda_0=\Lambda_1$,
would prove, mistakenly, that $S_{reg}^{00}=0$.

\subsection{Evaluation in Minkowski space}

Due to the simple two dimensional structure of the Schwinger model, it is an easy task to compute the vacuum polarization tensor, 
Eq.(\ref{vacpol}), directly in the Minkowski space, avoiding the Wick rotation.
 The convergent term $\Pi_{fin}^{\mu\nu}(p)$ is given, in the connection limit, by Eq.(\ref{finite}). For
the divergent term, $S^{\mu\nu}$, let us rewrite Eq.(\ref{PiInfFin}) as
\begin{equation}
S^{\mu\nu} = 2ie^2 \Big\{ \intk \frac{2k^{\mu}k^{\nu}}{(k^2-\mu^2)^2} - g^{\mu\nu} \intk \frac{1}{k^2-\mu^2} - \mu^2 g^{\mu\nu}\intk \frac{1}{(k^2-\mu^2)^2} \Big\}. \label{Smunu0}
\end{equation}
As before, let us introduce two different cutoffs to the integrals in $k_0$ and $k_1$. We obtain
\begin{eqnarray}
S_{reg}^{00}(p) &=& 2ie^2 \Big\{ \intSC \frac{2k_0^2}{(k_0^2-k_1^2-\mu^2)^2} - \intSC \frac{1}{k_0^2-k_1^2-\mu^2} \nonumber \\
&& - \mu^2 \intSC \frac{1}{(k_0^2-k_1^2-\mu^2)^2} \Big\}, \label{Pi00}
\end{eqnarray}
Solving the integral in $k_0$, we find
\begin{eqnarray}
S_{reg}^{00}(p) &=& i\frac{e^2}{\pi} \Big\{ \intkum \frac{2\Lambda_0}{k_1^2+\mu^2-\Lambda_0^2} - \intkum \frac{\mu^2 \Lambda_0}{(k_1^2+\mu^2)(k_1^2+\mu^2-\Lambda_0^2)} \nonumber \\
&& + \mu^2 \intkum \frac{1}{(k_1^2+\mu^2)^{3/2}} \arctan \Big( \frac{\Lambda_0}{\sqrt{-k_1^2-\mu^2}} \Big) \Big\}. \label{intk0}
\end{eqnarray}
The evaluation of the two first integrals in the right hand side of Eq.(\ref{intk0}) is straightforward. The last integral results
in a series of generalized hypergeometric functions. However, if it is evaluated for $\Lambda_1^2>>\mu^2$ (as $\Lambda_1^2 \to \infty$ and
$\mu^2$ is arbitrary this limit is exact) we obtain
\begin{eqnarray}
 S_{reg}^{00}(p) &=& i\frac{e^2}{2\pi^2} \Big\{ \frac{4\Lambda_0}{\sqrt{\mu^2-\Lambda_0^2}} \arctan \Big( \frac{\Lambda_1}{\sqrt{\mu^2-\Lambda_0^2}} \Big) 
+\frac{2\mu}{\Lambda_0}\arctan\Big(\frac{\Lambda_1}{\mu}\Big)
-\frac{2\mu}{\sqrt{\mu^2-\Lambda_0^2}} \arctan \Big( \frac{\Lambda_1}{\sqrt{\mu^2-\Lambda_0^2}} \Big) \nonumber \\
&& 
+\mu^2 \Big( \frac{\Lambda_1^2+\Lambda_0^2}{\Lambda_1^2 \Lambda_0^2} {\text{ arctanh}} \Big( \frac{\Lambda_0}{\Lambda_1} \Big) + \frac{1}{\Lambda_0\Lambda_1} \Big)
\Big\}. \label{integrated}
%
\end{eqnarray}
Except for the first term on the right hand side of Eq.(\ref{integrated}) all other terms vanish on taking the limits $\Lambda_0 \to \infty$ and
$\Lambda_1 \to \infty$. We are left with
\begin{equation}
S^{00} = \lim_{\Lambda_0, \Lambda_1 \to \infty} S_{reg}^{00} = i\frac{e^2}{2\pi^2} \lim_{\Lambda_0, \Lambda_1 \to \infty}  \frac{4\Lambda_0}{\sqrt{\mu^2-\Lambda_0^2}} \arctan \Big( \frac{\Lambda_1}{\sqrt{\mu^2-\Lambda_0^2}} \Big). \label{indetSC}
\end{equation}
Similar result is obtained for $S^{11}$.

Eq.(\ref{indetSC}) displays the indeterminacy on the vacuum polarization tensor in the sharp cutoff framework. Its result depends on the
path employed on the evaluation of the limits. For example, if one chooses to take $\Lambda_0 \to \infty$ as the first limit, and then take $\Lambda_1 \to \infty$,
it gives $S^{\mu\nu}=0$, and the gauge invariance of Eq.(\ref{vacpol}) is lost. 
Conversely, if one takes first $\Lambda_1 \to \infty$ and then $\Lambda_0 \to \infty$, one obtains $S^{\mu\nu}=\frac{e^2}{\pi} g^{\mu\nu}$.
This result shows that, in the second path, the symmetric integration trick is not valid, since the evaluation of Eq.(\ref{Pi00}) results in
a finite, non null, value. It also depends on the order of integration on $k_0$ and $k_1$: if we had integrated first in $k_1$ and
then in $k_0$, we should find
\begin{equation}
S^{00} = \lim_{\Lambda_0, \Lambda_1 \to \infty} S_{reg}^{00} = i\frac{e^2}{2\pi^2} \lim_{\Lambda_0, \Lambda_1 \to \infty}  \frac{4\Lambda_1}{\sqrt{\mu^2+\Lambda_1^2}} {\text{arctanh}} \Big( \frac{\Lambda_0}{\sqrt{\mu^2+\Lambda_1^2}} \Big).
\end{equation}
This is another face of the ill-defined character of the divergent integrals on Eq.(\ref{Smunu0}). It introduces two sources of
arbitraryness, one of them related with the mathematical indeterminacy of Eq.(\ref{indetSC}), when $\Lambda_0$ and $\Lambda_1$ approaches infinity,
and the other, associated with the non commutativity of the integrations on $k_0$ and $k_1$.

The choice of different cutoffs for each dimension can result in a loss of covariance, with a 
particular choice of the two-momentum cutoff $\Lambda^\mu=(\Lambda^0,\Lambda^1)$ breaking the Lorentz symmetry.
Covariance must then be reestablished {\it a posteriori}. Physically, in order to obtain Lorentz symmetric results, 
we should restrict ourselves to those paths that produce $S^{\mu\nu}=g^{\mu\nu} \alpha$. 
This is in the spirit of the covariant sharp cutoff regularization procedure. Nevertheless, we can see from Eq.(\ref{indetSC}) that
the result $S^{00} = 0$ is not unique - it is possible to choose an specific path to take the limits
in order to obtain $\lim_{\Lambda^0,\Lambda^1 \to \infty}\frac{\Lambda^\mu \Lambda^\nu}{\Lambda^2} = g^{\mu\nu}$.

For instance, it is also possible to choose an integration path that allows the obtaining of Lorentz covariance and gauge invariance - 
by evaluating each integral in the right hand side of Eq.(\ref{Wick}) directly in Minkowski space, integrating first on the variable
with the highest power on the integrand (i.e, integrating the first integral in Eq.(\ref{Wick}) in $k_0$ and the second one in $k_1$), we get
\begin{equation}
S^{\mu\nu} = i g^{\mu\nu} \frac{e^2}{\pi^2} \lim_{\Lambda_0, \Lambda_1 \to \infty} \Big\{ \frac{\Lambda_0}{\sqrt{\Lambda_0^2-\mu^2}} {\text{ arctanh}} \Big( \frac{\Lambda_1}{\sqrt{\Lambda_0^2-\mu^2}} \Big) 
+ \frac{\Lambda_1}{\sqrt{\Lambda_1^2+\mu^2}} {\text{ arctanh}} \Big( \frac{\Lambda_0}{\sqrt{\Lambda_1^2+\mu^2}} \Big) \Big\}, \label{SmunuSC}
\end{equation}
symmetric in the exchange $\Lambda_0 \leftrightarrow \Lambda_1$ for $\mu^2=0$. As one can see, the result is also undetermined. 
Thus, taking the limits $\Lambda_1 \to \infty$ and $\Lambda_0 \to \infty$ in this order, one obtains
\begin{equation}
S^{\mu\nu} = \frac{e^2}{2\pi} g^{\mu\nu},
\end{equation}
and 
\begin{equation}
\Pi^{\mu\nu}(p) = \frac{e^2}{\pi}\Big(g^{\mu\nu} - \frac{p^{\mu}p^{\nu}}{p^2} \Big),
\end{equation}
the expected gauge invariant result.

Finally, it is worth to stress that the tracelessness property of $S^{\mu\nu}$ is violated, in the sharp cutoff procedure, 
by the non commutativity of the two operations - the evaluation of the trace and the performing of the connection limits - on 
the surface term $\frac{\Lambda^\mu \Lambda^\nu}{\Lambda^2}$. In fact, taking the $\Lambda^\mu \to \infty$ limit properly,
we get
\begin{equation}
g_{\mu\nu} \lim_{\Lambda^0,\Lambda^1 \to \infty} \frac{\Lambda^\mu \Lambda^\nu}{\Lambda^2} = g_{\mu\nu} g^{\mu\nu} = 2
\end{equation}
or
\begin{equation}
\lim_{\Lambda^0,\Lambda^1 \to \infty} g_{\mu\nu} \frac{\Lambda^\mu \Lambda^\nu}{\Lambda^2} = \lim_{\Lambda^0,\Lambda^1 \to \infty} \frac{\Lambda^2}{\Lambda^2} = 1.
\end{equation}

\section{Pauli-Villars}

It is interesting to evaluate how gauge invariance is obtained in the framework of the Pauli-Villars regularization procedure.
To be consistent, one should start from the Schwinger Model with massive fermions, and then take the zero mass limit.
Introducing a set of $N$ Pauli-Villars fields with masses $\{M_i\}$ and couplings $\{C_i\}$, the vacuum polarization tensor is written as
\begin{equation}
\Pi^{\mu\nu}(p)=ie^2 \sum_{i=0}^{N} C_i \intk Tr \Big\{ \gamma^{\mu} \frac{\slash{k}+M_i}{k^2-M_i^2} \gamma^{\nu} \frac{\slash{k}+\slash{p}+M_i}{(k+p)^2-M_i^2} \Big\}, \label{vacpolPV}
\end{equation}
with $M_0 \to 0$, $C_0=1$ and $M_i \to \infty$ for $i \ne 0$. The coupling constants should be chosen in order to eliminate the (superficial) divergences on Eq.(\ref{vacpolPV}), i.e., we must have
\begin{equation}
\sum_{i=0}^{N} C_i=0. \label{C_i}
\end{equation}
One can see, from Eq.(\ref{vacpolPV}), that the vacuum polarization tensor in the Pauli-Villars procedure is not traceless due to the
presence of the masses $M_i$.
Evaluating the Dirac traces in Eq.(\ref{vacpolPV}) and introducing one Feynman parameter, one obtains
\begin{equation}
\Pi^{\mu\nu}(p)=ie^2 \sum_{i=0}^{N} C_i \Big\{ 2 S^{\mu\nu} - \frac{i}{2\pi} (2p^\mu p^\nu - g^{\mu\nu}p^2) \int_0^1 dx \frac{x(1-x)}{M_i^2 - p^2 x(1-x)} + \frac{i}{2\pi} g^{\mu\nu} M_i^2 \int_0^1 dx \frac{1}{M_i^2 - p^2 x(1-x)} \Big\},
\end{equation}
where $S^{\mu\nu}$ is now given by Eq.(\ref{PiInfFin}) with $M_i^2$ replacing $\mu^2$. After evaluating the $x$ integrals,
we get
\begin{equation}
\Pi^{\mu\nu}(p)= ie^2 \sum_{i=0}^{N} C_i \Big\{ 2S^{\mu\nu}_{\infty} + \frac{i}{\pi} \Big( \frac{p^\mu p^\nu}{p^2} - \frac{g^{\mu\nu}}{2} \Big) 
- \frac{i}{\pi} \Delta_i \Big( \frac{p^\mu p^\nu}{p^2} - g^{\mu\nu}  \Big) \Big\}, \label{vacpolPV1}
\end{equation}
with
\begin{equation}
\Delta_i=\frac{\xi_i}{\sqrt{(\xi_i-1)}} \arctan(\frac{1}{\sqrt{(\xi_i-1)}}) \label{Delta_i}
\end{equation}
and $\xi_i=\frac{4M_i^2}{p^2}$. The terms proportional to $\Delta_i$ in Eq.(\ref{vacpolPV1}) are absent in the other regularization
procedures we analyzed here, since $\lim_{M_i \to 0} \Delta_i = 0$. However, in the Pauli-Villars regularization, these are the only terms
that do not vanish after we apply the condition (\ref{C_i}). All the other terms vanish, since they do no depend on $M_i$, including
that do not vanish after we apply the condition (\ref{C_i}). All the other terms vanish, since they do no depend on $M_i$, including
the surface term $S^{\mu\nu}$. As one can see from Eq.(\ref{vacpolPV1}), these terms are gauge invariant. We thus are left with
\begin{equation}
\Pi^{\mu\nu}(p)= \frac{e^2}{\pi} \Big( \frac{p^\mu p^\nu}{p^2} - g^{\mu\nu}  \Big) \sum_{i=0}^{N} C_i \Delta_i. 
\end{equation}
From Eq.(\ref{Delta_i}) we obtain $\lim_{M_0 \to 0} \Delta_0 = 0$ and $\lim_{M_i \to \infty} \Delta_i = 1$. Together
with $\sum_{i=1}^N C_i = -C_0 = -1$, we get
\begin{equation}
\Pi^{\mu\nu}(p)=\frac{e^2}{\pi} \Big( g^{\mu\nu} - \frac{p^\mu p^\nu}{p^2}  \Big), \label{PimunuPV}
\end{equation}
the expected gauge invariant result. It is interesting to note that the mechanism that generates the gauge invariant result
in the Pauli-Villars regularization procedure is completely distinct from the dimensional and sharp cutoff cases. In fact,
the surface term that generates the indeterminacy is eliminated in this procedure by the condition (\ref{C_i}). It is also
an interesting point the fact that the final result, Eq.(\ref{PimunuPV}), does not depend on the number of Pauli-Villars fields
introduced in order to regulate the vacuum polarization tensor.


\section{The Chiral Schwinger Model}

We turn now to the analysis of the photon mass generation in the context of the Chiral Schwinger Model (CSM). The vector to fermions
coupling in the CSM is replaced by a chiral interaction, resulting in
\begin{equation}
L=i\bar{\psi}\partial\psi-e\bar{\psi} (1+\gamma_5) \gamma^{\mu}A_{\mu}\psi - \frac{1}{4}(F_{\mu\nu})^2, 
\end{equation}
with $\gamma_5=\gamma^0 \gamma^1$.

It is a well known result that the model can be evaluated exactly, and the RPA series of the vacuum polarization
tensor can be summed up resulting in a massive propagator with mass \cite{Jackiw:1984zi}
\begin{equation}
m^2=\frac{e^2}{\pi} \frac{a^2}{a-1}, \label{mass}
\end{equation}
where $a$ is an undetermined parameter associated with the regularization ambiguous term $S^{\mu\nu}$. In fact,
the evaluation of the vacuum polarization tensor in the CSM results in
\begin{equation}
\Pi^{\mu\nu}_{C}(p)= S^{\mu\nu} + \frac{2e^2}{\pi}g^{\mu\nu} - \frac{e^2}{\pi} (g^{\mu\rho}+\varepsilon^{\mu\rho}) \frac{p^\rho p^\sigma}{p^2} (g^{\sigma\nu}-\varepsilon^{\sigma\nu}),
\end{equation}
with Eq.(\ref{mass}) being obtained with $S^{\mu\nu}$ parametrized as $S^{\mu\nu}=(a-2) g^{\mu\nu} \frac{e^2}{\pi}$. This result reproduces
the well known fact that it is not possible to preserve gauge invariance in the chiral Schwinger model. It also displays the 
anomalous non conservation of the axial current \cite{BaetaScarpelli:2001ix}. It is worth to stress here that, as before, 
employing one of the regularization schemes 
previously analyzed in the vector Schwinger model will give us the undetermined result for $S^{\mu\nu}$, except, of course,
for the Pauli-Villars procedure. In order to get an unitary theory for the chiral Schwinger model, one has to have $a>1$ \cite{Jackiw:1984zi},
otherwise the theory would present tachyons, as can be easily observed from Eq.(\ref{mass}). This can be easily obtained in
the two approaches displayed on sections III and IV: in dimensional regularization, from Eq.(\ref{Piomegaalpha4}) one can see that $a>1$ implies in
$\delta=b \epsilon$, with $b>-1$. In sharp cutoff, it is possible to choose different paths to take the connection limits that
produce covariant results and interpolate between the two limits obtained in section IV.B, $S^{\mu\nu}=0$ and $S^{\mu\nu}=\frac{e^2}{\pi} g^{\mu\nu}$.
It can be easily shown that the condition $a>1$ is fullfilled by any path that produces $S^{\mu\nu} = \lambda g^{\mu\nu}$
with $\lambda > - \frac{e^2}{\pi}$.
Finally, in the context of the Pauli-Villars regularization scheme, the chiral Schwinger model was analyzed before in Ref. \cite{Xue:1991ge}.

\section{Conclusions}

In summary, we analyzed the appearance of a mathematical indeterminacy in several regularization schemes in the Schwinger and
chiral Schwinger models. We have shown that the use of regularization schemes with a set of two parameters ($\alpha$ and $\omega$
for dimensional regularization and $\Lambda_0$ and $\Lambda_1$ for sharp cutoff regularizations) allows the appearance of
an indeterminacy, in the mathematical sense, in the evaluatiuon of the vacuum polarization tensor. We also investigated why,
on preserving gauge invariance of the vector Schwinger model, the traceless of the vacuum polarization tensor is lost:
in dimensional regularization, it is not a property of the vacuum polarization in $2\omega$ dimensions, and in sharp cutoff, the non commutativity of
the connection limit with the trace operation is responsible for the lost of the tracelessness.

We also investigated why Pauli-Villars can reproduce gauge invariance in the vector Schwinger model, but concluded that this
procedure removes the essential ambiguity of the vacuum polarization tensor. Finally, the application
of the two parameters regularizations allows to obtain the undetermined character of the vacuum polarization tensor
and of the photon induced mass in the context of the chiral Schwinger model.

The extension of the present results for other models and theories at higher dimensions, for example for QCD in 4D, is
straigthforward. 

\section{Acknowledgments}
This research was supported by CAPES-Brazil. We are grateful to Dr. H. Caldas for usefull suggestions.


\end{document}